\documentclass[aps,floatfix,twocolumn]{revtex4}
\usepackage{graphicx}
\usepackage{graphics,epsfig}
\usepackage{amssymb}
\newcommand{\ETAL}{{et al.}}
\newcommand{\be}{\begin{equation}}
\newcommand{\ee}{\end{equation}}
\newcommand{\ba}{\begin{eqnarray}}
\newcommand{\ea}{\end{eqnarray}}
\newcommand{\bc}{}

\begin{document}
\title{Measuring the geometry of the Universe in the presence 
of isocurvature modes}
\author{J.~Dunkley$^{1}$}
\author{M.~Bucher$^{2}$}
\author{P.~G.~Ferreira$^{1,4}$}
\author{K.~Moodley$^{3,4}$}
\author{C.~Skordis$^{1}$}
\affiliation{
$^1$Astrophysics, University of Oxford, Denys Wilkinson Building, Keble Road, Oxford OX1 3RH, UK\\
$^2$Laboratoire de Physique Th\'eorique, Universit\'e Paris XI, 91405 
Orsay, France\\
$^3$Astrophysics \& Cosmology Research Unit and %
School of Mathematical \& Statistical Sciences, University of KwaZulu-Natal, Durban,
4041, South Africa\\
$^{4}$African Institute for Mathematical Sciences, 
6-8 Melrose Road, Muizenberg 7945, South Africa
}
\date{July 20, 2005}
\begin{abstract}
The Cosmic Microwave Background (CMB) anisotropy constrains
the geometry of the Universe because the positions of the acoustic 
peaks of the angular power spectrum depend strongly on the 
curvature of underlying three-dimensional space. 
In this Letter we exploit current observations to determine  
the spatial geometry of the Universe in the presence of isocurvature modes. 
Previous analyses have always assumed that the cosmological
perturbations were initially adiabatic.
{\it A priori} one might expect that allowing additional
isocurvature modes would substantially degrade the constraints on the 
curvature of the Universe. We find, however, that if one considers
 additional data sets, the geometry remains well constrained.
When the most general isocurvature perturbation is allowed,
the CMB alone can only poorly constrain 
the geometry to $\Omega_0=1.6\pm0.3$. Including 
large-scale structure (LSS) data one obtains
$\Omega_0=1.07\pm0.03,$ and  
$\Omega_0=1.06\pm0.02$ when supplemented by the
Hubble Space Telescope (HST) Key Project determination of $H_0$
and SNIa data. 
\end{abstract}
\maketitle

\begin{figure}[t!]
\epsfig{file=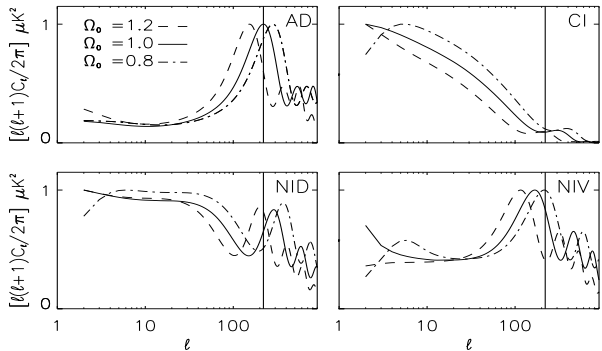,width=8.5cm,height=5.cm}
\vskip -0.15in
\caption{ 
The CMB temperature spectra for the adiabatic and  
isocurvature modes {\sc ad, ci, nid, niv} for $\Omega _0=0.8, 1.0, 1.2,$
where  
$\omega_b=0.022$, $\omega_d=0.12$, $\tau=0.1$, $n_s=1$, $h=0.7.$ 
\label{four_spectra} }
\vskip -0.25in
\end{figure}

One of the most striking successes of observational cosmology over 
the past few years has been the constraint of possible departures of the 
geometry of our universe from a spatially flat (Euclidean) geometry. Such
a departure from spatial flatness may be characterized by the dimensionless 
quantity 
$\Omega_K,$ related by the Friedmann-Robertson-Walker equations
to the fractional energy density (with respect to the
critical density) $\Omega_0$ of the Universe by the relation 
$\Omega_K=1-\Omega_0$. 
If $\Omega_K<0,$ the Universe has a three-dimensional spherical geometry,
if $\Omega_K>0,$ a hyperbolic geometry, and if $\Omega_K=0,$ 
a Euclidean (flat) geometry.
When a structure of a given physical size is viewed from a cosmological
distance, its angular size depends sensitively on the spatial geometry.
Different spatial geometries yield different laws of perspective.
Such differences may be exploited to obtain a measurement of $\Omega_K$
using the angular power spectrum of the CMB anisotropy \cite{omega_theory}. 
In the standard Big Bang thermal history, the Universe, previously ionized,
recombined into neutral atoms at a redshift $z_{rec}\approx 1080.$
At this moment the photon mean free path increased precipitously, 
and since then the CMB photons have travelled freely toward 
us with virtually no rescattering. Prior to recombination the previously
tightly coupled plasma had undergone acoustic oscillations, 
manifested in the CMB spectrum today by the so-called Doppler peaks.
At recombination these oscillations were 
characterized by a physical scale $d^*_{rec},$ the size
of the sound horizon at recombination.
In order to convert $d^*_{rec}$ into an angular scale $\theta ^*,$ another
length scale $d_{AD}$ is required. Let $d_{LS}$ \cite{eqn} be the diameter of
the last scattering surface expressed in terms of present
day co-moving units. In a flat geometry, it follows that
$\theta ^*=(z_{rec}+1)d^*_{rec}/d_{LS}.$ However,
in a curved geometry, $d_{LS}$ must be replaced with the 
apparent angular diameter distance $d_{AD},$ defined as
\ba
d_{AD}&=&{\Omega _K}^{-1/2}{H_0}^{-1}\sinh [{\Omega _K}^{1/2}H_0d_{LS}]
\cr
&=&d_{LS}\cdot [1+(1/6){\Omega _K}(H_0d_{LS})^2+\ldots ]
\ea
to account for the (de)focusing of rays by 
the nonzero spatial curvature \cite{eqn}.\\ 
\indent
The detection of the acoustic peak in the CMB by a number 
of experiments \cite{cmb_1stpeak} and
more recently with the Wilkinson Microwave Anisotropy Probe (WMAP) 
\cite{wmapdata} has been used to constrain the spatial curvature 
of the Universe. These
analyses, however, assumed that
the primordial perturbations were adiabatic---that is,  
at very early times the universe was governed by a common, spatially 
uniform equation of state and all the components 
contributing to the stress-energy shared a common 
peculiar velocity field. When this assumption is relaxed
other modes, the so-called isocurvature modes, arise, 
and their presence generically alters the positions of the Doppler
peaks used to determine the spatial curvature.
It is of interest to learn whether the constraints on $\Omega _K$
weaken significantly when the assumption of adiabaticity is 
relaxed. Previous studies have projected that without 
assuming adiabaticity the ability to determine
$\Omega _K$ with the WMAP data would be severely 
compromised \cite{bmt_2002}. In this Letter we examine this question and
find that $\Omega _K$ is poorly constrained by the 
CMB data alone; however, when LSS structure data
is included, one is able to impose stringent
constraints on $\Omega _K.$\\
The data sets considered here 
include
the CMB anisotropy data and the SDSS determination of
the galaxy power spectrum, as well as 
the determination of the Hubble constant 
and the luminosity-redshift relation for Type Ia Supernovae.
The fluctuations of the CMB 
are characterized by the angular power spectrum 
$C_\ell$ where 
$\langle \vert \Delta T_{lm}\vert ^2\rangle =\ell(\ell+1)C_\ell/2\pi ,$  
which may be transformed into an angle using
$\ell\approx \pi /\theta $. For a cosmology with adiabatic initial
cosmological perturbations (and other initial conditions as well), 
the angular CMB power spectrum $C_l$ exhibits
a series of peaks and troughs whose positions scale according
to the above determination of $\theta ^*.$
For a Euclidean universe of an age consistent 
with current measurements of the Hubble constant \cite{HST} and 
adiabatic initial 
conditions, the first so-called Doppler peak is situated 
at $\ell\approx 220,$ 
as observed by WMAP \cite{wmapdata}, leading to 
constraints, assuming adiabatic initial conditions 
of $\Omega_0 = 1.09\pm^{0.06}_{0.13}$ \cite{sloan_param}, or 
$0.98<\Omega_0 <1.08$ at 95\% confidence, assuming a Hubble constant of 
$h>0.5$ \cite{wmap_param}. The geometry was further constrained to 
$\Omega_0 = 1.02\pm0.02$  \cite{wmap_param,sloan_param} by 
including data from the 2dFGRS or SDSS  
galaxy surveys \cite{2df,sloan}, supernovae redshift-luminosity relation
\cite{SN} and the HST key project determination of $h$ \cite{HST}.\\
We now consider more general initial conditions 
where isocurvature modes and possible correlations among themselves and with 
the 
adiabatic mode are also allowed. Under this framework
four additional perturbation modes are possible: the CDM (CI), baryon 
(BI), neutrino density (NID) and neutrino velocity (NIV) isocurvature 
modes, discussed in detail in \cite{iso_theory,bdfms}.
Constraints on these modes have been presented 
in \cite{lit_iso_postWMAP,bdfms} for flat cosmologies.
Each of these modes predicts distinguishable CMB power spectra templates 
shown in Fig. \ref{four_spectra}, except for the BI and CI modes 
which have almost identical spectra. An admixture of new modes can 
shift the peak positions and weaken the curvature constraint.\\
We consider a family of cosmological models characterized by five free
parameters, the  
curvature
$\Omega_k$, the baryon density parameter $\omega _b=\Omega_bh^2$, the 
cold dark 
matter density $\omega _d=\Omega_dh^2$, a cosmological 
constant $\Omega_{\Lambda}$ and 
the optical depth to recombination $\tau$ subject to the constraint 
$\tau<0.3$, 
all which are assigned uniform priors. 
We impose the weak priors $0<\Omega_b<1$ and $0<\Omega_d<3$ and 
consider neither a varying dark energy equation of state nor 
neutrino masses. With four distinguishable scalar perturbation modes, 
labelled by 
$i,j={\mbox{\small{AD}}},{\mbox{\small{CI}}},{\mbox{\small{NID}}},
{\mbox{\small{NIV}}}$, 
the initial perturbation power spectrum consists of  
ten symmetric matrix elements 
$\Delta^{ij}(k)$, 
parametrized as 
$\Delta^{ij}(k)\propto (k/k_0)^{\frac{(n_i+n_j)}{2}-1}$, 
introducing a scalar spectral index $n_i$ for each mode and a
pivot scale taken as $k_0=0.05h/{\mbox {Mpc}}$. 
The angular power spectrum of temperature and polarization 
anisotropies corresponds to the quadratic observable
\be 
C_{\ell AB}^{ij}=\int _0^\infty \frac{dk}{k}\Delta^{ij}(k)
\Theta_{\ell A}^i(k)\Theta_{\ell B}^j(k)
\ee
where $\Theta_{\ell A}^i(k)$ is the radiation transfer function for 
the mode $i$ and observable $A=T$ or $E$, $T$ indicating the temperature
anisotropy and $E$ the electric polarization. 
We compute the ten components of the 
theoretical CMB temperature and polarization spectra 
$C_\ell^{ij}$ and 
matter power spectra $P^{ij}_m(k)$ for each correlation using the 
publicly available {\sc camb} package \cite{camb}. 
To generate all ten sets of spectra for $l<1600$ and $P_m(k)$ up to 
$k=0.7h/{\mbox {MPc}}$ takes about 20 sec
for a flat model and 70--100 sec
for curved models using two 3GHz processors.
We modified the code to sum the spectra with varying amplitudes 
using the methods described in \cite{bdfms}, 
including nine relative amplitudes and one overall amplitude 
(or one amplitude for adiabatic models). The 
ten $z_{ij}$ coefficients quantify the relative contribution 
to the power in the CMB by each correlation, 
with $\sum_{i,j=1}^4z_{ij}^2=1$. $z_{ad}$ has $i,j={\mbox{\small{AD}}}$.

\begin{figure}[t!]
\vskip -0.1in
\epsfig{file=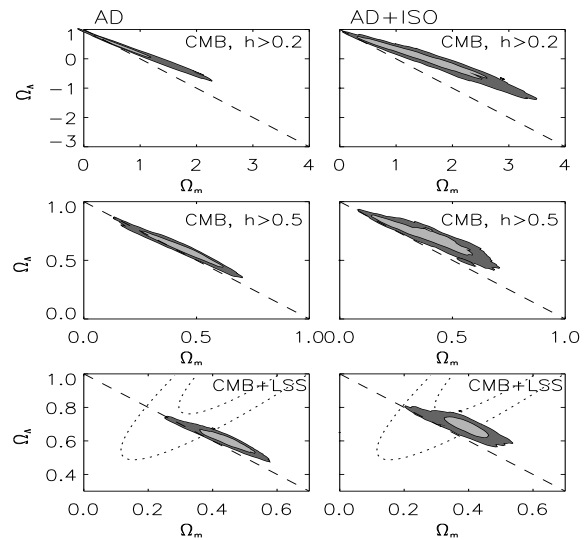,width=8cm,height=7.7cm} 
\vskip -0.25in
\caption{ 
Marginalized 68\% (pale grey) and 95\% (dark grey) constraints on 
$\Omega_m$ and $\Omega_\Lambda$ for adiabatic (left, $\mbox{\small{AD}}$) 
and mixed 
initial condition models (right, $\mbox{\small{AD+ISO}}$). The top two rows use CMB data, 
the bottom row includes LSS data ($h>0.5$) and shows constraints from SN Ia 
data alone (dotted). 
\label{twod_lambda} }
\vskip -0.27in
\end{figure}
\begin{figure}[t!]
\vskip -0.1in
\epsfig{file=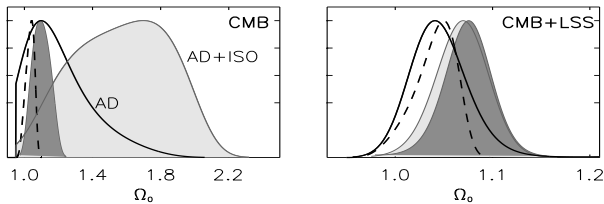,width=8.5cm,height=3.4cm} 
\vskip -0.15in
\caption{ 
Marginalized posterior distributions for $\Omega_0$ for adiabatic
models ({\sc ad},unshaded) and mixed initial conditions with 
common spectral indices ({\sc ad+iso}, shaded). 
The dark shaded and dashed curves show results for $h>0.5$, otherwise
$h>0.2$.
\label{oned_omega} }
\vskip -0.3in
\end{figure}

We compare models to the WMAP temperature and 
temperature-polarization CMB data covering $2<\ell<800$ \cite{wmapdata} 
using the likelihood function in \cite{verde}, together with 
small-scale CMB data  covering $800<\ell<1600$ \cite{tegmarkcomp,acbar}. To 
increase computational speed we do not use $\ell>1600$ 
data. We
also add the SDSS galaxy power spectrum data 
$P_g(k)$ in the
linear regime $k<0.2h/{\mbox {Mpc}}$ \cite{sloan}. We include 
a bias parameter $b$ such that $P_g(k)=b^2 P_m(k)$
and use the non-linear approximation for $P_m(k)$ in \cite{halofit} 
implemented in {\sc camb}. Finally, the analysis is repeated including
the likelihood as a function of
$\Omega_m$ and $\Omega_\Lambda$ inferred from SN Ia redshift-luminosity 
measurements
\cite{SN}, and also the Gaussian prior on $h$ of 
$0.72\pm0.08$ obtained by HST \cite{HST}.
The posterior distributions are sampled using the Markov Chain 
Monte Carlo 
methods described 
in \cite{dunkley}. We derive $\Omega_0=1-\Omega_k$ and the isocurvature 
fraction $f_{iso}=z_{iso}/(z_{iso}+z_{ad})$, where $z_{iso}^2=1-z_{ad}^2$, 
quoting marginalized median and 68\% confidence intervals. 
We test the software by reproducing results for flat models obtained 
with the DASh code \cite{dash,bdfms}.\\ 
\begin{figure}
\vskip -0.2in
\epsfig{file=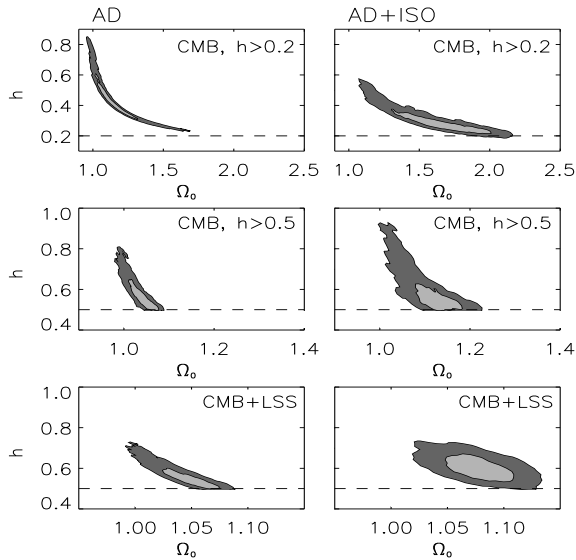,width=8cm,height=8cm}
\vskip -0.25in
\caption{ 
Constraints on the Hubble constant $h$ and the fractional energy density 
$\Omega_0$, with panels as in Fig. \ref{twod_lambda}.
\label{twod_hubble} }
\vskip -0.15in
\end{figure}
\begin{table}[t!]
\begin{center}
\begin{tabular}{lcc}
\hline\hline
Dataset &  $h$ & $\Omega_0$ \\
\hline
{\sc cmb} & $>0.2$  & $1.2\pm_{0.1}^{0.2}$ \\
{\sc cmb}& $>0.5$ & $1.03\pm0.03$ \\
{\sc cmb + lss} & $>0.2$ & $1.04\pm0.03$ \\
{\sc cmb + lss} & $>0.5$ & $1.04\pm0.02$\\
{\sc cmb + lss + sni}a & $0.72\pm0.08$ & $1.01\pm^{0.02}_{0.01}$\\
\hline
\end{tabular}
\end{center}
\vskip -0.15in
\caption{Median and $68\%$ confidence intervals for the 
fractional energy density $\Omega_0$ for adiabatic models 
with minimum Hubble constant $h$ or Gaussian HST prior $0.72\pm0.08$.}
\label{table1}
\vskip -0.2in
\end{table}
We first explored the determination of the spatial curvature
allowing only the adiabatic models, characterised by
seven parameters (eight with LSS), and including various combinations
of data sets. With only the CMB data and the 
weak prior $0.2<h<1$, the 
curvature is poorly constrained as shown in 
Table \ref{table1}. The CMB provides a 
good measure of the angular diameter distance 
$d_{AD}=(z_{rec}+1)d^*_{rec}/\theta ^*$, but 
a degeneracy exists between the curvature and 
the conformal distance to last-scattering $d_{LS}$\cite{geom}, as
illustrated in the top-left panels of 
Figs.~\ref{twod_lambda} and \ref{twod_hubble}.
$d_{LS}$ increases with decreasing $h$ or
$\Omega_\Lambda$ and with increasing $\Omega_m.$ 
The constraint on the geometry is considerably tightened when 
$h>0.5$ is imposed. The degeneracy is also broken by the LSS data.
The shape of the power spectrum constrains
$\Omega_mh,$ ruling out the closed models with high $\Omega_m$ allowed by 
the CMB data, which instead constrains $\Omega_mh^2$. 
Similarly supernova  
luminosity-redshift measurements exclude models with high $\Omega_m$ and 
low $\Omega_\Lambda$, as shown in Fig.~\ref{twod_lambda}. By including 
CMB, LSS, SNIa 
and HST data combined, the geometry is strongly constrained to
$\Omega_0=1.01\pm^{0.02}_{0.01}$, in agreement with \cite{wmap_param,sloan_param}. 
\begin{figure}[t!]
\vskip -0.2in
\epsfig{file=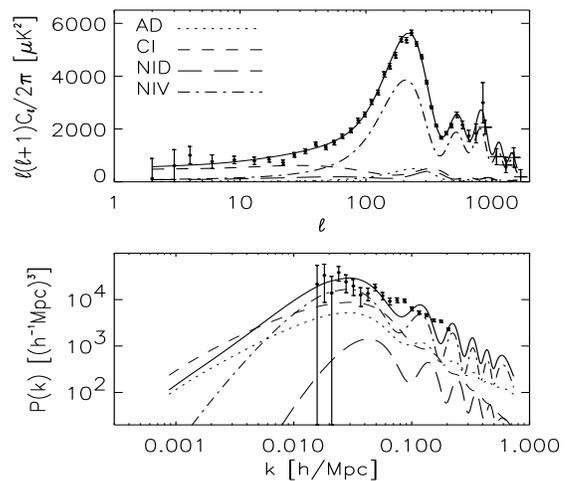,width=8cm,height=7.4cm} 
\vskip -0.25in
\caption{ 
CMB temperature (top) and galaxy power spectra (bottom) for a 
curved cosmological model dominated by isocurvature with 
$f_{iso} = 90\%$, $\Omega_0=1.9$. It fits current CMB data well 
(TT and TE), but does not fit LSS data and is inconsistent with 
BBN and HST observations ($\omega_b=0.08$, $h=0.3$). 
\label{niv_spec}}
\vskip -0.2in
\end{figure}
\begin{table}[t!]
\vspace{-.05in}
\begin{center}
\begin{tabular}{lcccc}
\hline\hline
Dataset & $h$ &$\Omega_0$ & $\langle f_{iso}\rangle $ \\
\hline
{\sc cmb } & $>0.2$ &$1.6\pm0.3$ & $0.6$ \\
{\sc cmb } & $>0.5$ &$1.10\pm0.05$ & $0.6$ \\
{\sc cmb + lss} & $>0.2$ & $1.07\pm0.03$ & $0.5$\\
{\sc cmb + lss } & $>0.5$ & $1.07\pm0.03$ & $0.5$ \\
{\sc cmb + lss +sni}a & $0.72\pm0.08$ & $1.06\pm0.02$ & $0.54$ \\
{\sc cmb + lss +bbn} & $>0.5$ & $1.06\pm0.03$ & $0.31$ \\
\hline
\end{tabular}
\end{center}
\vskip -0.1in
\caption{Median and $68\%$ confidence intervals for $\Omega_0$ 
and median isocurvature fraction $f_{iso}$ for models with mixed initial conditions.}
\label{table2}
\vskip -0.2in
\end{table}
We now consider how the measurements are affected when
the full range of correlated isocurvature and adiabatic 
initial conditions are admitted.
Initially the spectral indices of the modes are fixed as
a single parameter, 
giving rise to a family of models described by sixteen parameters 
(seventeen with LSS).
Figs.~\ref{twod_lambda},~\ref{twod_hubble} 
and \ref{oned_omega} and 
Table \ref{table2} demonstrate that the determination of 
the geometry of the Universe is
significantly degraded by the inclusion of these additional degrees
of freedom. 
As in the adiabatic cases, where 
$\Omega_0=1.2\pm^{0.2}_{0.1}$ results from the CMB data alone, 
the data favor a closed Universe
with $\Omega _0=1.6\pm0.3$ when isocurvature is allowed, with the priors
$h>0.2$ and $\Omega_d<3$
constraints cutting off larger curvatures. The extremely closed models,
such as the one in Fig \ref{niv_spec}, are dominated by up to 
90\% isocurvature (at 2$\sigma$), and exhibit low $h$, 
high $\Omega_m$ and low $\Omega_\Lambda$, extending the degenerate direction
observed for the adiabatic models.
Many models dominated by the NIV mode
require a high baryon content ($\omega_b\approx 0.06$),
so the acoustic peak positions are shifted to smaller scales
by lowering the sound speed at recombination.
This effect contributes to the exclusion of 
isocurvature-dominated extreme open models. 
Closed models dominated by the NID or CI modes do not fit the CMB data, 
particularly at large scales. 

Additional data constraining $d_{LS}$ 
improve the curvature constraints.
The prior $h>0.5$ alone suffices to exclude highly closed
models.
Alternatively, the LSS data 
constrains the allowed range to $\Omega_0=1.07\pm0.04$
in two ways. Firstly, its shape 
constrains $\Omega_mh$ and rules out high $\Omega_m$ models, 
since in most models the matter power spectrum is dominated by
the adiabatic mode. 
Secondly, the LSS data rules out models dominated by 
NIV, which predict significant baryon oscillations. 
The supernova and HST data further 
constrain $d_{LS}$. The combination of all datasets gives 
$\Omega_0=1.06\pm0.02$, with $\langle f_{iso}\rangle =0.54$. 
The distribution for $f_{iso},$ however, is very broad and
does not appear to disfavor the adiabatic models. 
Within this class of models the data prefer 
higher baryonic densities 
($\omega_b=0.04\pm0.02$ for CMB+LSS data)
than consistent with nucleosynthesis 
measurements. We find that adding the 
Gaussian prior  $\omega _b=0.022\pm0.002$ 
from recent BBN measurements \cite{bbn} to 
the CMB+LSS analysis with the prior $h>0.5$ 
reduces the  
isocurvature contribution by just under a half 
and yields $\Omega_0 =1.06\pm0.03.$
We also repeated the analysis for an enlarged space of models
where each of the isocurvature modes is given an independent tilt
within the range $0<n_i<2$.
The resulting twenty-dimensional model constrained by
CMB+LSS with the $h>0.5$ prior 
gives $\Omega_0 = 1.08 \pm0.03$ 
and the following
constraints on the spectral indices:
$n_{\mbox{\tiny{AD}}}= 1.1\pm0.2$, $n_{\mbox{\tiny{CI}}}= 1.1\pm0.5$, $n_{\mbox{\tiny{NID}}}= 1.4\pm0.3$, 
$n_{\mbox{\tiny{NIV}}}= 1.1\pm0.3$, 
with isocurvature fraction $\langle f_{iso}\rangle =0.5$. The
region of high probability density, however, includes models with
virtually no isocurvature. We do not find that the models with
isocurvature modes offer a better fit than the adiabatic models,
beyond what may generically be expected as a result of enlarging
the number of degrees of freedom of the theoretical model. 
This result is analogous to 
the flat case considered in \cite{bdfms}. \\
In this letter, we have substantially strengthened
the evidence for an almost flat Universe, with the data suggesting
a slight preference for a mildly closed Universe. 
In considering a very general class of initial conditions,
we have removed the assumption of adiabaticity and hence on specific
models of the early universe such as inflation. In principle this could
have completely removed the ability to constrain the geometry of the
Universe with current data sets, as was shown in \cite{bmt_2002}. Indeed
we find with this wider
range of parameters a behavior similar to what has already been noted for
purely adiabatic initial conditions: that {\it the CMB alone does not
suffice to constrain the geometry of the Universe.}
Once one includes 
additional information (constraining $h$ or $\Omega_m h$), it is
possible to pin down the geometry with appreciable precision. 
Consequently, for 
very general initial conditions where
perturbations are imprinted 
at early times and subsequently evolved through gravitational collapse, we 
conclude that the Universe is very nearly flat.\\
{\it Acknowledgements}:  
We thank O.~Lahav and R.~Trotta for discussions. 
J.D. and C.S are supported by PPARC.
%

\end{document}